\documentclass[usenatbib]{mn2e}
\usepackage{amsmath,amssymb}
\usepackage{graphicx}
\usepackage{hyperref}
\newcommand{\fig}[1]{Fig. \ref{#1}}
\newcommand{\refsection}[1]{Section \ref{#1}}
\newcommand{\reftable}[1]{Table \ref{#1}}
\newcommand{\refeq}[1]{Eq. (\ref{#1})}

\begin{document}
\title{Hydrodynamic Effects in the Symmetron and $f(R)$-gravity Models}

\author[Amir Hammami et al.]
{Amir ~Hammami,$^1$
Claudio ~Llinares,$^1$ David F. ~Mota,$^1$ and Hans A. Winther, $^2$ \\
$^1$Institute of Theoretical Astrophysics, University of Oslo, P.O. Box 1029 Blindern, N-0315 Oslo, Norway\\
$^2$Astrophysics, University of Oxford, DWB, Keble Road, Oxford, OX1 3RH, UK\\}

\maketitle

\begin{abstract}
In this paper we present the first results from implementing two scalar-tensor modified gravity theories, the symmetron and the Hu-Sawicki $f(R)$-gravity model, into a hydrodynamic N-body code with dark matter particles and a baryonic ideal gas. The study is a continuation of previous work where the symmetron and $f(R)$ have been successfully implemented in the RAMSES code, but for dark matter only. By running simulations, we show that the deviation from $\Lambda$CDM in these models for the gas density profiles are significantly lower than the dark matter equivalents. When it comes to the matter power-spectrum we find that hydrodynamic simulations agree very well with dark matter only simulations as long as we consider scales larger than $k\sim 0.5$ h/Mpc. In general the effects of modified gravity on the baryonic gas is found to not always mirror the effects it has on the dark matter. The largest signature is found when considering temperature profiles. We find that the gas temperatures in the modified gravity model studied here show deviations, when compared to $\Lambda$CDM, that can be a factor of a few larger than the deviations found in density profiles and power spectra.
\end{abstract}

\section{Introduction}

The current standard model of cosmology is the $\Lambda$CDM model, and even though it is remarkably accurate in many aspects, it still has several problems that remains to be solved \citep{LCDMprob1}. Furthermore, $\Lambda$CDM has no proper motivation for adding the $\Lambda$ constant other than the fact that it gives rise to the acceleration of the expansion of the Universe \citep{LCDMprob2}. This can also be explained by introducing a new fluid known as dark energy \citep{darkenergy} to the matter content of the Universe. Alternatively, the acceleration of the Universe might be a signal that gravity is modified on the largest scales. Modifying gravity is usually done by modifying the Einstein-Hilbert Lagrangian density $\mathcal{L}_{\rm EH}=R$ by replacing it with a more general function including terms of higher-order in derivatives of the metric ($R^2$, $R_{\mu\nu}R^{\mu\nu}$, $R_{\alpha\beta\mu\nu}R^{\alpha\beta\mu\nu}\ldots$) or by introducing new dynamical degrees of freedom, such as scalar fields \citep{Capozziello}, coupled to the matter sector \citep{brook}.

General Relativity have been thoroughly tested in the lab \citep{labexp,labexp2} and in the solar-system and no deviation have so far been found \citep{cwillreview,solarsyst}. This places strong constrains on any model that seeks to modify gravity and for such models to have any cosmological signatures, apart from modifying the background expansion history, a screening mechanism is required. A screening mechanism \citep{screen1,screen2} is a way of suppressing the effects of modifications of gravity in high density regions, compared to the critical density, such as on Earth and in the solar-system. In this paper, we will take a look at the symmetron  \citep{SymmetronPaper} and $f(R)$-gravity models \citep{husawicki}, which are but two of numerous scalar-tensor field theories that possess such a screening mechanism.

With the ever increasing number of theoretical modified gravity models it is important to find ways to compare these models to observations in order to exclude the ones that are not viable. Local gravity experiments typically gives constrains on modified gravity models that translates into signatures being in the non-linear regime of structure formation. Therefore one of the ways to find useful observables is to use N-body simulations. So far the majority of the cosmological N-body community that works with modified gravity has focused solely on simulations made with collisionless dark matter \citep{fofrnbodychicago,2008arXiv0809.2899L,Li1,Li2,Li3,Barreira,MG-Gadget,nbodydgp,nbodycham,nbodyvainst,nbodysymmnqs,symmfofrredshift, ISIS}. However in doing so we neglect a wealth of physics, after all what is observed are photons emitted from luminous matter. To go beyond dark matter and try to describe the plethora of baryonic effects that take place in out Universe is very challenging. Over the last decade, several codes have been written that combine N-body simulations with hydrodynamics \citep{Ramses,Gadget,MethodologyPaper,Illustris, RyOo} and include recipes for handling star formation and feedback processes.

So far there has only been, to our knowledge, two attempts at combining modified gravity with hydrodynamics in N-body codes \citep{MG-Gadget,nrtwo}. In this paper we show the first results from a hydrodynamic N-body code with the symmetron model, and also reproduce and expand upon the same $f(R)$-gravity theory as presented in \citet{MG-Gadget} and \citet{nrtwo}. The code that this paper is based on is a slight modification of ISIS \citep{ISIS}, which in turn is a modification of the RAMSES hydrodynamic N-body code \citep{Ramses}. This paper will only concern itself of the modifications made in order to combine the modified gravity part of the N-body code with the hydrodynamic part of RAMSES.  For more on the implementation of the scalar fields and other technicalities we refer the reader to \citet{ISIS}. The aim of this paper is to study the effects of modified gravity on the baryonic gas, compared to the effects that modified gravity has on the dark matter physics. For this purpose we will focus on simple observables such as the matter power spectrum and density and temperature profiles for both the dark matter and gas components separately.

The structure of this paper is as follows. In \refsection{scaltenstheory} we give an introduction to scalar-tensor theories of gravity and take a look at how the equations differ from those of standard gravity. \refsection{code_implem} briefly details the code implementations and the run parameters of our simulations. We show our results for density profiles in \refsection{profile_results}, for temperature profiles in \refsection{temp_prof_results} and for power spectra in \refsection{power_results}. We finish the paper with a short discussion in \refsection{discuss}.

\section{Scalar-tensor theories of gravity}
\label{scaltenstheory}

We are interested in scalar-tensor theories that can are defined by the following action \citep{Sotiriou:2006hs,2003CQGra..20.4503F}:
\begin{align}
 S &= \int d^4x\sqrt{-g}\left[\frac{R}{2}M_{pl}^2 - \frac{1}{2}\partial^i\psi\partial_i\psi - V(\psi)\right] \label{Sym_action} \\
 &\quad+ S_m(\tilde{g}_{\mu\nu},\tilde{\Psi}_i),  \nonumber
\end{align}
where $g$ is the determinant of the metric tensor, $g_{\mu\nu}$ in the Einstein frame, which is related to the Jordan frame metric tensor $\tilde{g}_{\mu\nu}$ by a conformal factor $A(\psi)$,
\begin{align}\label{conftrans}
 \tilde{g}_{\mu\nu} = A^2(\psi)g_{\mu\nu},
\end{align}
and $R$ is the Ricci scalar. The conformal factor satisfy $A \simeq 1$ for all the models we consider in this paper and we will use this approximation throughout. For more on these frames and the transformations between them and possible errors see \citet{Faraoni:1998qx} and \citet{BrownHammami}.

Varying the action with respect to the scalar field and the metric gives us the scalar field equation of motion and the stress energy-tensor respectively,
\begin{align}
 \Box\psi = V'(\psi) - A'(\psi) T^{(m)}, \label{Sym_EoM}
\end{align}
where $T^{(m)}$ is the trace of the stress energy tensor, ${T^{(m)}=g^{\mu\nu}T^{(m)}_{\mu\nu}}$,  and
  \begin{align}
 T_{\mu\nu} &= A(\psi)T_{\mu\nu}^{(m)} + T_{\mu\nu}^{(\psi)}\nonumber\\
&= A(\psi)\left[(P+\rho)u_{\mu}u_{\nu} + Pg_{\mu\nu}\right]  \label{Sym_stress} \\
&\quad+ \nabla_{\mu}\psi\nabla_{\nu}\psi - g_{\mu\nu}\left(\frac{1}{2}\partial^i\psi\partial_i\psi + V(\psi)\right). \nonumber 
\end{align}

Note that the total stress-energy tensor is covariantly conserved, while the scalar field component itself is not
\[\nabla^{\nu}T_{\mu\nu}^{(\psi)}\neq0.\] 

We will now briefly go through the derivation of the fluid equations we get by using basic principles on the above action, and show what form they take for the symmetron model and the $f(R)$-theory.

\subsection{The fluid equations}
We can derive the fluid equations of scalar-tensor theories from basic principles. We use the general action \refeq{Sym_action} and stress-energy tensor \refeq{Sym_stress} and start by computing the covariant derivative of the stress-energy tensor, which is a conserved quantity \citep{Gravitation}
\begin{align}
\nabla^{\mu}T_{\mu\nu} &= 0 .
\end{align}
We work in the Newtonian Gauge,
\begin{align}
 ds^2 = -(1+2\Phi)dt^2 + a^2(1-2\Phi)\delta_{ij}dx^idx^j,
\end{align}
and from the conservation of the stress-energy tensor and by imposing the quasi-static limit \citep{2013PhRvL.110p1101L, quasistatic} for the scalar field, in which time-derivatives of the scalar field is ignored relative to spatial gradients, we find the fluid equations for scalar-tensor theories of gravity,
\begin{align}
&\frac{\partial \rho}{\partial t} + \nabla(v\rho)  +3H\rho = 0,\\
a^2(P+\rho)\Big[Hv &+ \frac{\partial v}{\partial t} + (v\cdot\nabla) v + \frac{1}{a^2}\nabla\Phi\Big] \\
+ \nabla P &+ \frac{A'(\psi)}{A(\psi)}\rho\nabla\psi = 0, \nonumber \\
\frac{\partial E}{\partial t} + 2HE &+ \frac{P}{\rho}\cdot\nabla v = - (v\cdot\nabla)\Phi - \frac{A'(\psi)}{A(\psi)} (v\cdot\nabla)\psi.
\end{align}

We now perform a change of variables by implementing a variation of the so-called super-comoving coordinates, introduced by \citet{SuperCom},
\begin{align}
 d\tilde{t} = a^{-2}dt,\qquad  \tilde{\rho} = a^3\rho,\qquad  \tilde{v} = a^2v, \\
 \tilde{\psi} = a\psi, \qquad \tilde{P} = a^5P,\qquad  \tilde{\Phi} = a^2\Phi, \qquad \tilde{E} = a^2E,
\end{align}
where the tildes represent quantities in the super-comoving coordinates. The purpose of these coordinates is to eliminate unwanted dependencies on the scale and Hubble factors. The equations then take the form\footnote{This results is reached by excluding terms of second order and assuming static pressure and fields.}
\begin{align}
\frac{\partial \tilde{\rho}}{\partial \tilde{t}} + \nabla(\tilde{v}\tilde{\rho}) &= 0,\label{final_cont}\\
\frac{\partial \tilde{v}}{\partial \tilde{t}} + (\tilde{v}\cdot\nabla)\tilde{v} = - \frac{1}{\tilde{\rho}}\nabla \tilde{P} &-\nabla\tilde{\Phi} - \frac{A'(\tilde{\psi})}{A(\tilde{\psi})}\nabla\tilde{\psi} \label{final_euler}, \\
\frac{\partial \tilde{E}}{\partial \tilde{t}} + \tilde{v}\cdot\nabla\tilde{E} + \frac{\tilde{P}}{\tilde{\rho}}\cdot\nabla \tilde{v} = - (\tilde{v}\cdot&\nabla)\tilde{\Phi} - \frac{A'(\tilde{\psi})}{A(\tilde{\psi})} \tilde{v}\cdot\nabla\tilde{\psi},\label{final_energy}
\end{align}
which are the equations we have implemented in the N-body code.

\subsection{Symmetron model}
The symmetron model was introduced by \citet{SymmetronPaper} as a new screening mechanism, similar to the one found in the Chameleon models \citep{Chameleons,motashaw}. It utilises a screening mechanism so that we recover General Relativity in regions of high density (such as the solar system where gravity has been very well tested) whereas we get an order one modification of gravity in low density regions. This is done by introducing a potential on the symmetry breaking form 
\begin{align}
V(\psi) = V_0-\frac{1}{2}\mu^2\psi^2 + \frac{1}{4}\lambda\psi^4,
\end{align}
where $\psi$ is the scalar field, $\mu$ is a mass scale and $\lambda$ a dimensionless parameter. 

The  coupling factor just mentioned is also chosen to be symmetric in the same manner as the potential 
\[A(\psi) = 1 + \frac{1}{2}\left(\frac{\psi}{M}\right)^2 \]
where $M$ is another mass scale.

From the field equation we have that the dynamics of the field is determined by an effective potential given by,
\begin{align}
V_{\rm eff}(\psi) = V_0 + \frac{1}{2}\left(\frac{\rho_m}{M^2}-\mu^2\right)\psi^2 + \frac{1}{4}\lambda\psi^4,
\end{align}
In regions of high density ($\rho_m \gg M^2\mu^2$) the field is driven towards the minimum $\psi = 0$, while in regions of low density we get a minimum at $\psi_0 = \pm\mu\sqrt{\frac{1}{\lambda}}$ for which the field will reside close to. The fifth-force (see below) is proportional to the local value of the scalar field so in high density regions $\psi \approx 0$ and it will be suppressed.

We want, however, to work with slightly other parameters, for which the physical interpretation is more clear, as presented in \citet{Winther}. This entails changing our free parameters $\mu$, $M$ and $\lambda$ to $\beta$, $\lambda_0$ and $a_{\rm SSB}$ like
\begin{align}
\label{sym_params}
 \beta  &= \frac{M_{pl}\psi_0}{M^2}, \\
a_{\rm SSB}^3 &= \frac{3H_0^2\Omega_m M_{pl}^2}{M^2\mu^2}, \\
\lambda_0^2 &= \frac{1}{2\mu^2}.
\end{align}
Now $\beta$ represents the strength of the scalar fifth-force (relative to the gravitational force), $a_{\rm SSB}$ is the expansion factor at the time of symmetry breaking, and is also related to the density at which the screening mechanism kicks in via the relation $\rho_{\rm SSB}=\Omega_{m0}\rho_{c0}a_{\rm SSB}^{-3}$, and $\lambda_0$ is the range of the scalar fifth-force in units of Mpc/h. Further the scalar field itself is replaced by a dimensionless scalar field $\chi$ by
\begin{align}
 \tilde{\psi} &= \psi_0\chi.
\end{align}
The equation of motion for this scalar field in the quasi-static limit is \citep{ISIS}
\begin{align}
 \nabla^2\chi = \frac{a^2}{2\lambda_0}\left[\left(\frac{a_{\rm SSB}}{a}\right)^3\frac{\rho_m}{\overline{\rho}_m} + \chi^3 - \chi\right].
\end{align}
Simulations beyond the static limit were presented in \citet{2013PhRvL.110p1101L, nbodysymmnqs}, finding only sub-percent differences between the static and non-static solutions. 

The fifth-force in the symmetron model takes the form
\begin{align}
F_{\psi} &= \frac{A'(\tilde{\psi})}{A(\tilde{\psi})}\nabla\tilde{\psi} = \frac{\tilde{\psi}}{M^2}\nabla\tilde{\psi} \nonumber \\
&= 6\Omega_m H_0^2\frac{(\beta\lambda_0)^2}{a_{\rm SSB}^3}\chi\nabla\chi,
\end{align}
which is how it is represented in our code. For more on the symmetron model we refer to \citet{SymmetronPaper}.

\subsection[nowarningplease]{$f(R)$-gravity}
All $f(R)$-gravity theories revolve around promoting the Ricci scalar in the Einstein-Hilbert action to a function of the Ricci scalar instead. In the N-body code the Hu-Sawicki $f(R)$ model \citep{husawicki} has been implemented and is this is the model we focus on in this paper. The action for $f(R)$-gravity is given by
\begin{align}
 S = \int\sqrt{-g}\left[\frac{R+f(R)}{16\pi G} + \mathcal{L}_m\right]d^4x,
\label{husawick}
\end{align}
and in the Hu-Sawicki model $f(R)$ has the form
\begin{align}
 f(R) = -m^2\frac{c_1(R/m^2)^n}{1 + c_2(R/m^2)^n},
\end{align}
where $n$, $c_1$ and $c_2$ are the free parameters and $m^2=H_0^2\Omega_{m0}$. We can reduce the number of free parameters from three to two by demanding that $c_1 = 6c_2\frac{\Omega_{\Lambda}}{\Omega_m}$ (to yield dark energy) as demonstrated in \citet{husawicki}. Further it is convenient to use a parameter $f_{R0}$ instead of $c_2$ \citep{ISIS}
 \begin{align}
  f_{R0} = -\frac{6n\Omega_{\Lambda}}{c_2\Omega_m}\left(\frac{\Omega_{\Lambda}}{3(\Omega_m + 4\Omega_{\Lambda})}\right)^{n+1}.
 \end{align}
By applying the conformal transformation \refeq{conftrans} to the action \refeq{husawick} using
\begin{align}
A(\psi) = e^{\frac{-\beta\psi}{M_{pl}}}
\end{align}
with $\beta = 1/\sqrt{6}$ we recover the general scalar-tensor theory action \refeq{Sym_action}. This model is then further characterised by the $f_R$ function
\begin{align}
\label{fRr}
 f_R = A^2(\psi)-1 \approx -\frac{2\beta\psi}{M_{pl}}.
\end{align}
In order to get the field-equation on a convenient form for a numerical implementation we must perform a change of variables. As justified in \citet{fofrnbodychicago} we introduce a substitution of the form
\begin{align}
 f_R = -a^2e^u,
\end{align}
combining this with \refeq{fRr} we see that this is essentially a substitution like
\begin{align}
 \psi = a^2\frac{M_{pl}}{2\beta}e^u \Rightarrow \nabla\psi = a^2\frac{M_{pl}}{2\beta}e^u\nabla u.
\end{align}
In \citet{fofrnbodychicago} it has been shown that the equation of motion for this scalar field is then
\begin{align}
 \nabla\cdot(e^u\nabla u) &= \Omega_maH_0^2\Bigg[(\tilde{\rho} - 1) +\left(1+4a^3\frac{\Omega_{\Lambda}}{\Omega_m}\right) \\
&\quad -a^3\left(1+4\frac{\Omega_{\Lambda}}{\Omega_m}\right)\left(a^2f_{R0}\right)^{\frac{1}{n+1}}e^{-\frac{u}{n+1}} \Bigg],\nonumber 
\end{align}
where $f_{R0}$ is related to the range of the scalar field via $\lambda_0 \propto 1/\sqrt{f_{R0}}$. 
The fifth-force is given by
\begin{align}
 F_{\psi} &= \frac{A'(\tilde{\psi})}{A(\tilde{\psi})}\nabla\tilde{\psi} = \frac{a^2\beta}{M_{pl}}\nabla\tilde{\psi} \nonumber \\
&= \frac{1}{2}e^u\nabla u,
\end{align}
which is how it is represented in our code. For more on $f(R)$-theories of gravity see the review by \citet{f(R)Theories}.

\section{Code Implementation and run parameters}
\label{code_implem}
Implementing scalar-tensor theories of gravity to the hydrodynamic part of the N-body code is rather straightforward, thanks to the scalar-tensor theories all giving contributions as a fifth-force and the fact that RAMSES, which ISIS is based on, has been widely used, thoroughly tested and optimised. Wherever the code normally works with the gravitational force, call it $F_{GR}$, we simply replace it with an effective force $F_{\rm eff}$ that includes the effects of modified gravity
\begin{align}
 F_{\rm eff}=F_{GR}+F_{\psi},
\end{align}
which is then naturally permeated throughout the code. For more on the implementation of the scalar field solver itself see \citet{ISIS}. In the next two subsections we present our initial conditions and other parameters we used in our runs. For both models we also made a run with standard gravity ($\Lambda$CDM) to use as a base of comparison. All initial conditions have been generated by using the Grafic code, which is a part of COSMICS \citep{Grafic}, based on the parameters described below.

\subsection{Parameters for the symmetron simulations}
The symmetron simulations were run using 1024 cores, $256^3$ dark matter particles, with a box width of $256$ Mpc/h and six levels of refinements. The background cosmology is a standard $\Lambda$CDM background with $h=0.65$, $\Omega_{\Lambda} = 0.65$, $\Omega_m=0.35$ and $\Omega_b = 0.05$.  
\begin{table}
 \begin{center}
\caption{Overview of the model parameters for the symmetron and $f(R)$ models.\label{tab:mgparam}}
  \begin{tabular}{lrrr}\hline
  Name  & $\quad\beta$ & $\quad a_{\rm SSB}$ & $\quad\lambda_{\psi}$ \\ \hline
  Sym\_A & \quad1.0 & \quad0.5 & \quad1.0 \\ 
  Sym\_B & \quad1.0 & \quad0.33 & \quad1.0 \\ 
  Sym\_C & \quad2.0 & \quad0.5 & \quad1.0 \\ 
  Sym\_D & \quad1.0 & \quad0.25 & \quad1.0 \\ \hline\hline
  Name   & $\quad f_{R0}$&$\quad n$& \\ \hline
  FofR04 & $\quad10^{-4}$&\quad1& \\ 
  FofR05 & $\quad10^{-5}$&\quad1& \\ 
  FofR06 & $\quad10^{-6}$&\quad1& \\ 
  \end{tabular}
 \end{center}
 
\end{table}

The symmetron parameters are presented in \reftable{tab:mgparam}. These parameters were chosen to focus on various aspects of the symmetron models. Varying $\beta$ changes the strength of the fifth-force while changing $a_{\rm SSB}$ changes the time of symmetry breaking (i.e. it is the scale-factor for which the fifth-force kicks in) and also the density criteria for the screening mechanism.

\subsection{Parameters for the f(R) simulations}
The $f(R)$ simulations were run using 1024 cores, $256^3$ dark matter particles with a box width of $200$ Mpc/h and eight levels of refinements. The background cosmology is a standard $\Lambda$CDM background with $h=0.70$, $\Omega_{\Lambda} = 0.727$, $\Omega_m=0.272$ and $\Omega_b = 0.045$.\footnote{These $\Lambda$CDM parameters were chosen to coincide with those of \citet{MG-Gadget} to enable cross-checking.}

The $f(R)$ parameters are presented in \reftable{tab:mgparam}. These parameters as briefly introduced above were chosen as they give the full range of effects found in the model: from almost no screening and large deviations from $\Lambda$CDM for FofR04 to much screening and small deviations from $\Lambda$CDM for FofR06. The main effect of changing $f_{R0}$ is to change the range over which the fifth-force is acting on and also the density threshold for screening.

\section{Density profiles}
\label{profile_results}
In this section, we present density profiles for multiple halos identified by using the the Rockstar code developed by \citet{rockstar} and incorporated into the YT-Project \citet{YT-Project}. We filter all the halos that have not reached a relaxed state. This is done by following the methods described in \citet{relax} and \citet{Shaw}, where we use relations between the kinetic- and potential energy and the surface pressure to determine if a halo is relaxed or not. We used the method presented by \citet{symmfofrredshift} to take into account the effects of modified gravity in the virialization state of the halos.  A halo is defined to be relaxed if $\left|\frac{2T - E_s}{U}\right| \leq 0.2$.  See \citet{symmfofrredshift} for more on this limit and its implications.

\begin{figure*}
Dark Matter \hspace{5.5cm}\;Gas\\
\vspace{-2mm}
        \centering
        \includegraphics[width=0.4\textwidth]{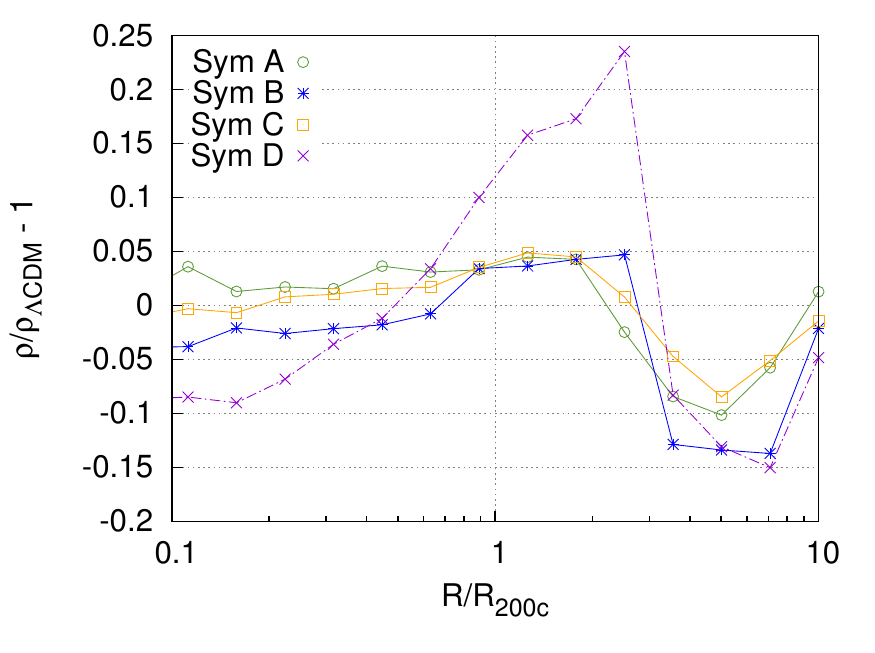}
	\vspace{-5 mm}
	\hspace{-5 mm}
        \includegraphics[width=0.4\textwidth]{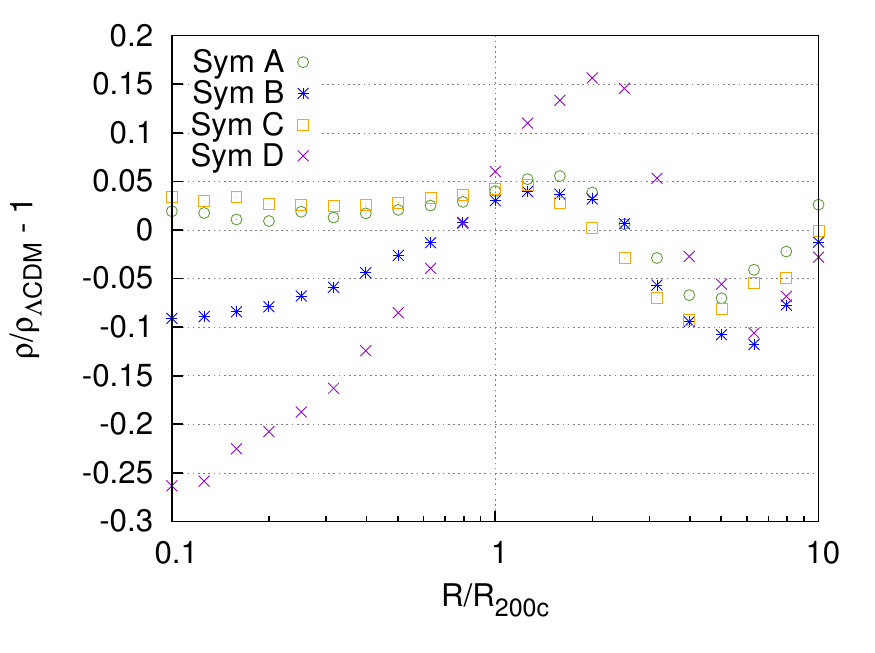}

        \includegraphics[width=0.4\textwidth]{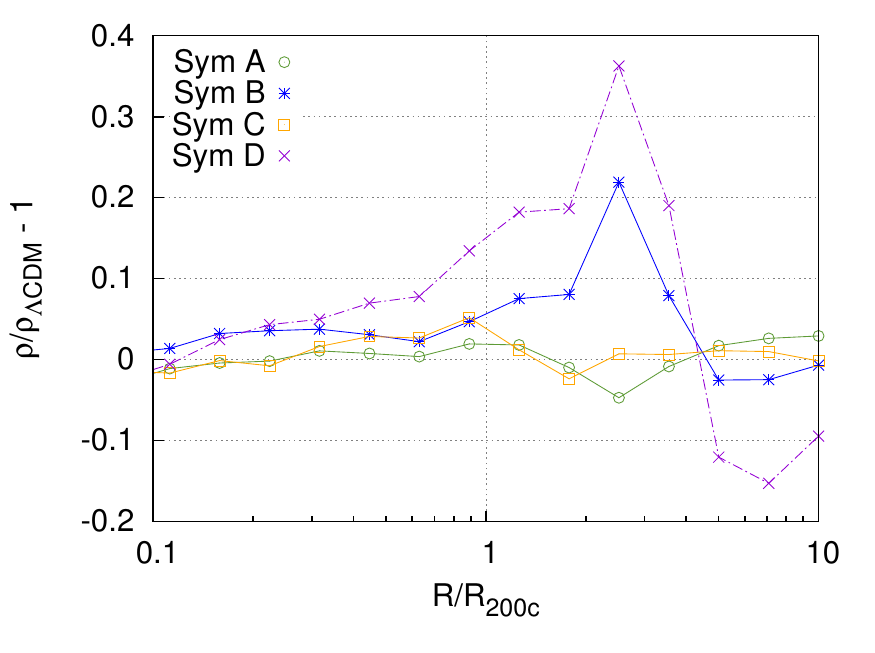}
	\vspace{-5 mm}
	\hspace{-5 mm}
        \includegraphics[width=0.4\textwidth]{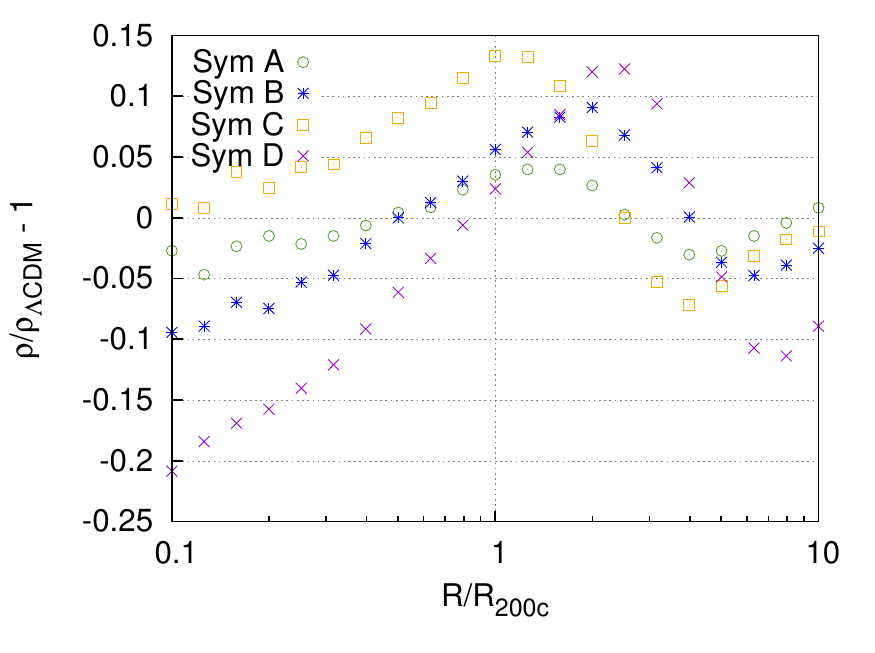}
	\vspace{-1 mm}
\caption{The top figures show the deviations from $\Lambda$CDM for our symmetron models for CDM and baryons respectively for all density profiles with a mass between $10^{14}$ $h^{-1}M_{\odot}$ and $5\times10^{14}$ $h^{-1}M_{\odot}$. The bottom figures show the same for for all density profiles with a mass between $10^{13}$ $h^{-1}M_{\odot}$ and $5\times10^{13}$ $h^{-1}M_{\odot}$.}
 \label{fig:Symmetron_density_profiles}
 \end{figure*}

We are also interested in seeing how the behaviour depends on the mass of the halos.  We will therefore study halos with masses in the ranges 1-5$\times10^{13} h^{-1}M_{\odot}$ and 1-5$\times10^{14}h^{-1}M_{\odot}$. The density profiles are calculated by binning dark matter particles and the baryonic gas density in annular bins for each halo, then averaging over all halos. We focus only on the present day epoch which corresponds to $z=0$. 

Our calculated density profiles are averages of all density profiles of the proper size, ranging from 10\% of the virialization radius, $r=0.1R_{\rm 200c}$, to ten times the virialization radius, $r=10R_{\rm 200c}$.  This range was chosen to properly catch all behaviours of the fifth-force on the dark matter and gas halos while also avoiding the inner regions of the halos where the resolution of our simulations is not sufficient.  

\begin{figure*}
Dark Matter \hspace{5.5cm}\;Gas\\
\vspace{-2mm}
        \centering
        \includegraphics[width=0.4\textwidth]{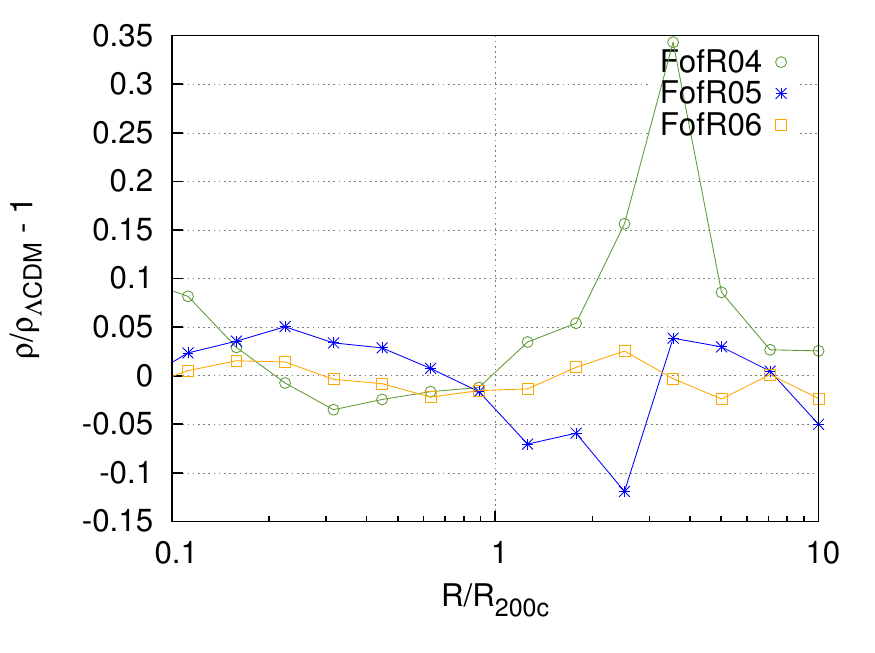}
	\vspace{-5 mm}
	\hspace{-5 mm}
        \includegraphics[width=0.4\textwidth]{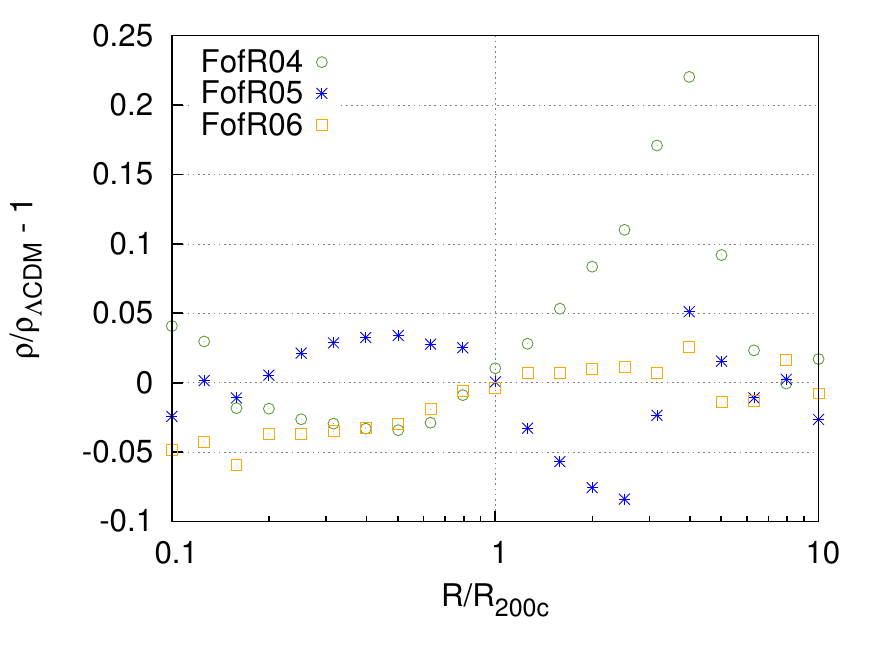}

        \includegraphics[width=0.4\textwidth]{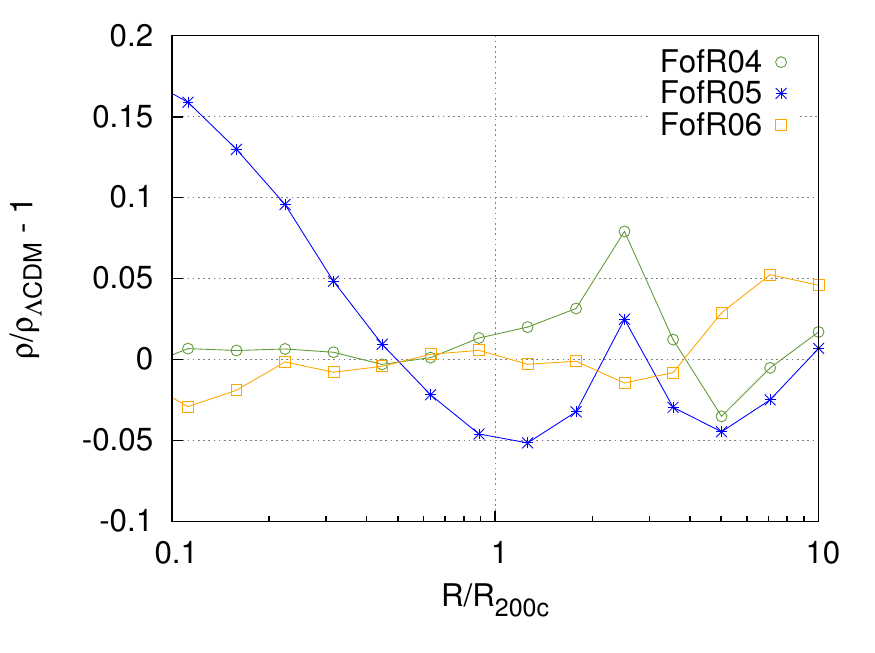}
	\vspace{-5 mm}
	\hspace{-5 mm}
        \includegraphics[width=0.4\textwidth]{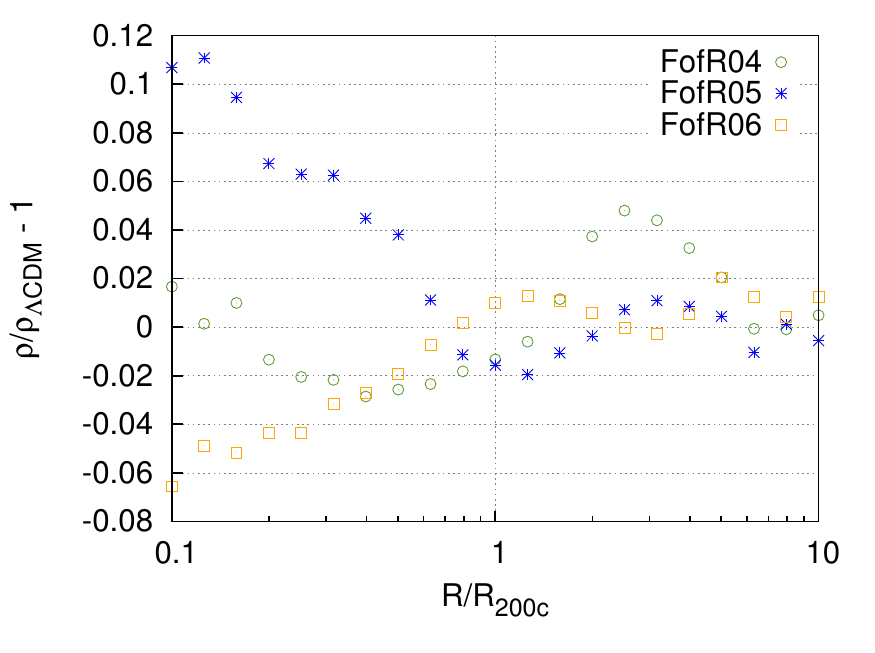}
	\vspace{-1 mm}
\caption{The top figures show the deviations from $\Lambda$CDM for our $f(R)$ models for CDM and baryons respectively for all density profiles with a mass between $10^{14}$ $h^{-1}M_{\odot}$ and $5\times 10^{14}$ $h^{-1}M_{\odot}$. The bottom figures show the same for for all density profiles with a mass between $10^{13}$ $h^{-1}M_{\odot}$ and $5\times 10^{13}$ $h^{-1}M_{\odot}$.}\label{fig:FofR_density_profiles}
\end{figure*}

In \fig{fig:Symmetron_density_profiles} we present the deviations from $\Lambda$CDM of the density profiles for the dark matter and the baryonic gas for the symmetron model, while in \fig{fig:FofR_density_profiles} we present the same figures for the $f(R)$-gravity model. 

The dark matter density profiles for modified gravity show in general stronger clustering at the outskirts of the halos than the inner regions. The additional force from the scalar field will increase the rate at which dark matter and the gas collapses towards the centre. When the density criteria is then met and the fifth-force is screened the dark matter particles will have gained such velocities that they overshoot the centre of the halo and does not cluster there. The particles that overshoot the centre will then end up in the outskirts as we see in the density profile deviations.

We observe that overall that the gas profiles behave closer to the $\Lambda$CDM counterpart then we find in the dark matter profiles. In previous works similar effects on the power spectra have been attributed to AGN feedback \citep{MG-Gadget}. In our simulations however we do not have these effects, as we chose to implement only the most rudimentary of baryonic components, a perfect fluid, clearly this effect is much more intrinsic to baryons than before assumed.
 
What we see here might be an environmental effect, from the dark matter, on the gas component. The dark matter collapses faster than the gas component, due to the CDM being collisionless while the gas components are hindered from collapsing due to friction and pressure as enforced by the Euler equations \refeq{final_cont} -\refeq{final_euler}, this means that the dark matter will cluster faster than the gas and reach higher densities earlier than the gas. The total density, dark matter and gas, will trigger the density criteria that turns on the screening mechanism before the baryonic gas has had a chance to collapse as much as the dark matter. In other words, dark matter cuts off the fifth-force before it has had the time to work on the gas component to the same effect as it does on the dark matter.

As seen in \reftable{tab:mgparam}, we remember that all our symmetron simulations have the same coupling strength and force range, but with varying symmetry breaking criteria. The exception is Sym C which has the same symmetry breaking criteria as Sym A, but with twice the coupling strength. Due to this we will temporarily ignore Sym C when studying the effects of the symmetron parameters.

We immediately see that the lower $a_{\rm SSB}$ is, meaning the fifth-force has been acting upon the Universe for a longer time and also screens at a higher density, the bigger the deviations from $\Lambda$CDM are in the extremities of the density profile plots, with the differences between the models being smallest near the virialization radius. One might na\"{i}vely expect that models with a lower $a_{\rm SSB}$ should break off from $\Lambda$CDM closer to the centre of the halo, where the density is higher, however it is clear from these profiles that the length of time that the fifth-force has been working on the particles is the stronger of the two effects.

Returning to Sym C, we note that the increased coupling strength amplifies the amplitude of Sym A, while also making it behave as if it had a slightly lower $a_{\rm SSB}$ than it actually has, in short it simply boosts the effects of the fifth-force. Looking at the low-mass figures we note that the profiles follow the same trend as before, but that the effects from the stronger coupling in Sym C is much stronger. This is due to the fact that the halos of this mass range have sizes that are approximately of the order of the force range and we get a stronger resonance. 

For the $f(R)$ case the main difference between the parameters is the range of the fifth-force, and the behaviour of these $f(R)$-gravity models have been discussed before in \citet{BajiLi}. What is worth noticing is that we observe the same behaviour when comparing the dark matter density profiles to the gas density profiles as we did in the symmetron case. The only exception is for the high mass halos in the FofR06 simulation, which is the model closest to $\Lambda$CDM, as is expected, in the dark matter case. However, the gas profiles show slightly lower clustering in the inner regions than what we have for the dark matter (similar to what we see for Sym B, but to a larger extent). 

\section{Temperature profiles}
\label{temp_prof_results}
\begin{figure*}
        \centering
	\vspace{-5 mm}
        \includegraphics[width=0.48\textwidth]{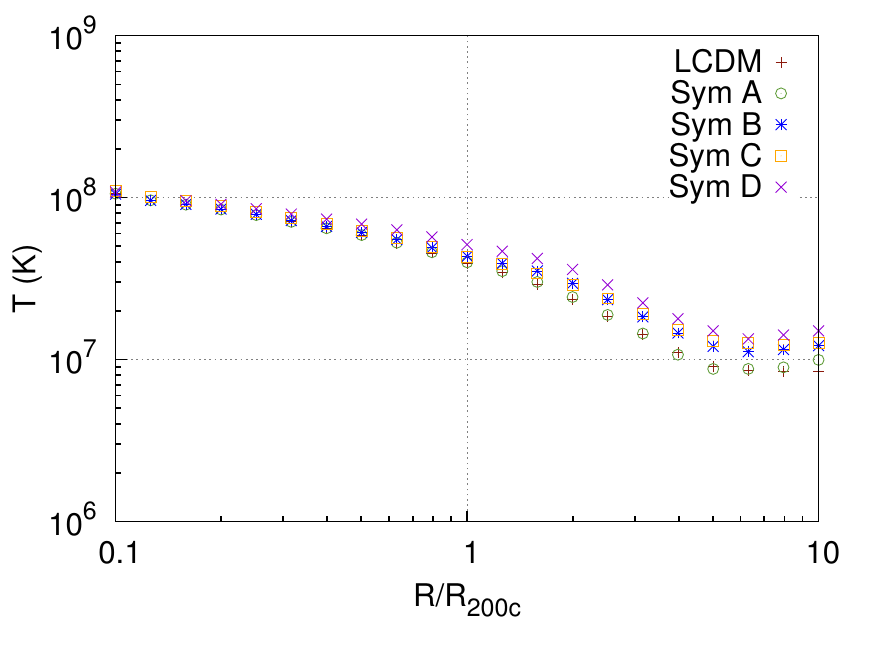}
	\vspace{-5 mm}
        \includegraphics[width=0.48\textwidth]{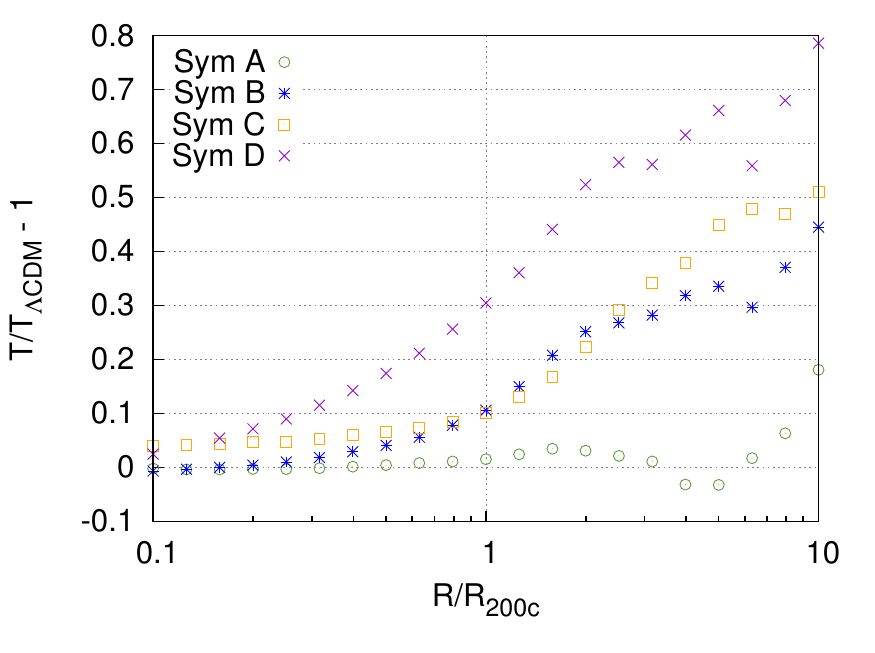}
	\vspace{-1 mm}
\caption{The left figure show the gas temperature profiles for $\Lambda$CDM and our symmetron models for all halos with a mass between $10^{14}$ $h^{-1}M_{\odot}$ and $5\times 10^{14}$ $h^{-1}M_{\odot}$, the right figure show the deviations from standard $\Lambda$CDM for the same models.}\label{fig:temperature_profiles_Sym}
\end{figure*}

\begin{figure*}
        \centering
        \includegraphics[width=0.48\textwidth]{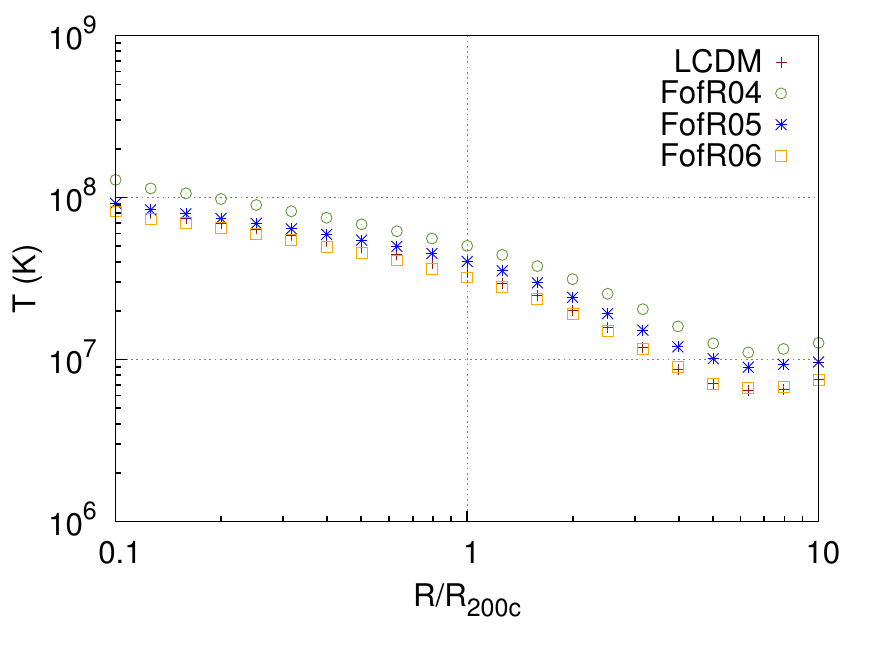}
        \includegraphics[width=0.48\textwidth]{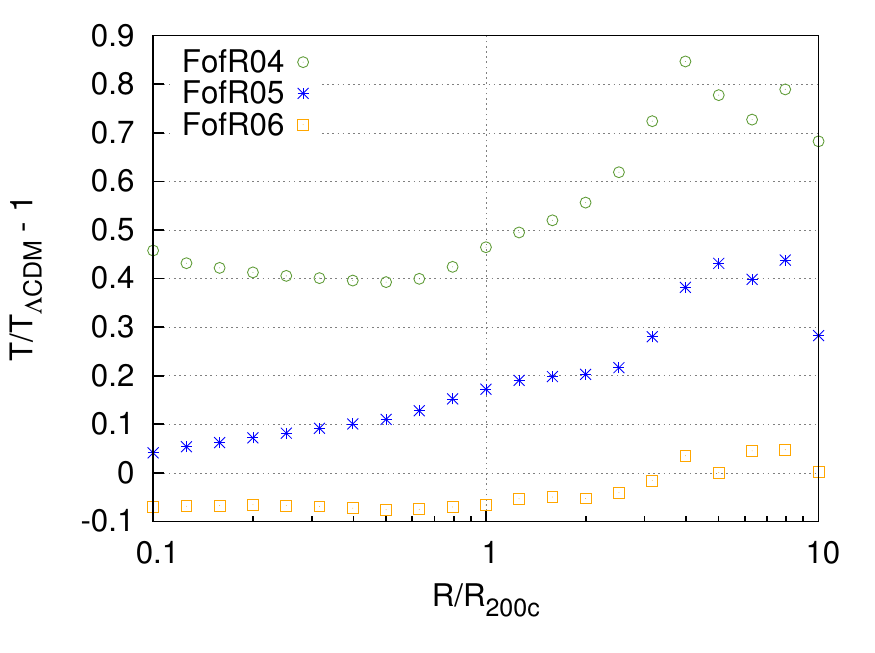}
	\vspace{-1 mm}
\caption{The left figure show the gas temperature profiles for $\Lambda$CDM and our $f(R)$  models for all halos with a mass between $10^{14}$ $h^{-1}M_{\odot}$ and $5\times 10^{14}$ $h^{-1}M_{\odot}$, the right figure show the deviations from standard $\Lambda$CDM for the same models.}\label{fig:temperature_profiles_FofR}
\end{figure*}

In this section we provide the temperature-profiles for the halos in our simulations. The motivation for showing this is that the temperature of the gas in our Universe is one of the direct, and easiest, observables that we can use to compare to our results, furthermore this is something that, to our knowledge, has never before been presented for modified gravity theories. To calculate the temperature from the simulations we utilise the ideal gas law,
\begin{align}
p = R_{s}\rho T,\label{temp_def}
\end{align}
where $p$ is the thermal pressure, $R_{s}=\frac{k_B}{m_H}$ is the specific gas constant and $p/\rho$ is provided to us from the simulation output\footnote{The profiles we show are mass-weighted, i.e. $T = \frac{\sum m_i T_i}{\sum m_i}$}. In \fig{fig:temperature_profiles_Sym} and \fig{fig:temperature_profiles_FofR} we show both the profiles and the deviations in the profiles with respect to $\Lambda$CDM for our symmetron and $f(R)$-gravity models respectively.

\begin{figure}
        \centering
        \includegraphics[width=0.48\textwidth]{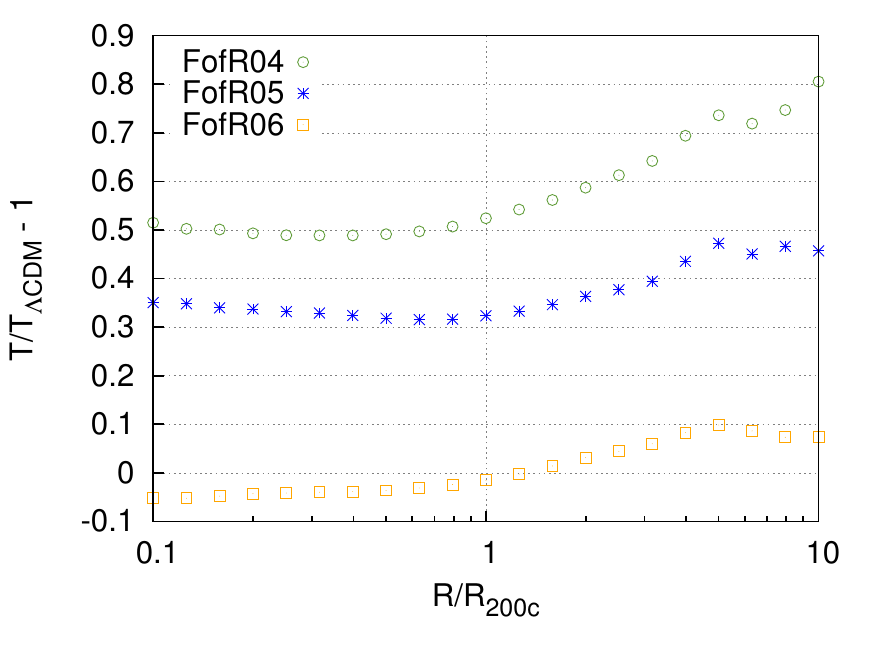}
	\vspace{-1 mm}

\caption{Gas temperature deviations from standard $\Lambda$CDM for our $f(R)$ models for halos with a mass between $10^{13}$ $h^{-1}M_{\odot}$ and $5\times10^{13}$ $h^{-1}M_{\odot}$.}\label{fig:temperature_profiles_FofR_lowmass}
\end{figure}

\begin{figure}
        \centering
        \includegraphics[width=0.48\textwidth]{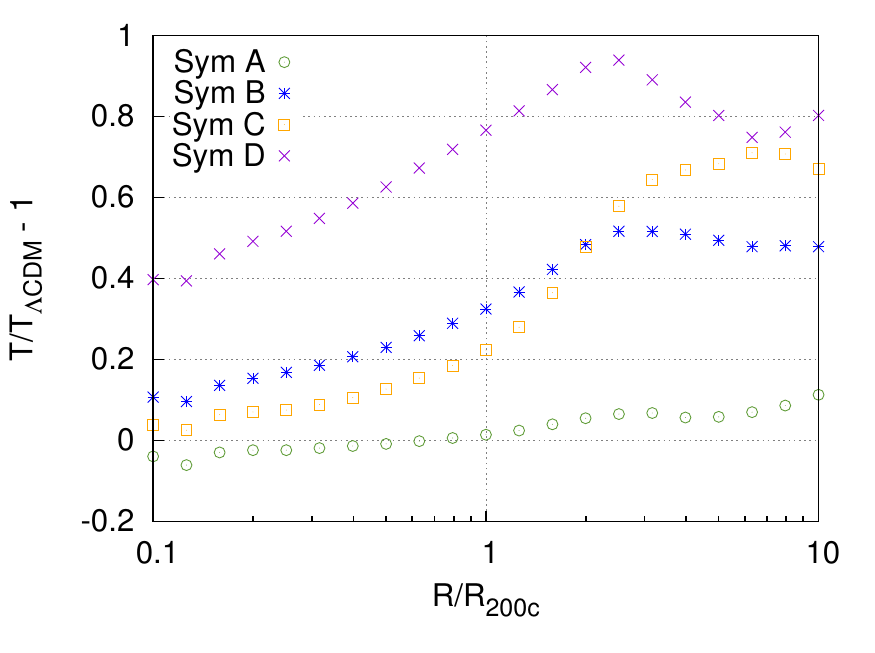}
	\vspace{-1 mm}
\caption{Gas temperature deviations from standard $\Lambda$CDM for our symmetron models for halos with a mass between $10^{13}$ $h^{-1}M_{\odot}$ and $5\times10^{13}$ $h^{-1}M_{\odot}$.}\label{fig:temperature_profiles_Sym_lowmass}
\end{figure}

For the models we have simulated the screening properties for (dark matter) NFW halos have been studied in \cite{2014arXiv1412.0066G} and these results are very useful to understand the results we find. 

The first thing to notice from the temperature profiles is that the effect of modified gravity can be larger than what we find in the density profiles and power-spectra. This is a similar kind of signature as has been found in the velocity field in modified gravity simulations previously \citep{2013MNRAS.428..743L} and not surprising as the temperature of the gas is closely related to how fast (and faster means more turbulent) the gas is moving.

The largest halos we study are so massive today that they have been screened (as modifications of gravity is a late time effect in our models) during most of their evolution. This is also reflected in the temperature profiles for FofR05, FofR06 and all the symmetron models where we see a close to zero deviation in the centre of the largest halos. The only exception is FofR04 where we see a deviation even in the center. As seen in Fig. 6 of \cite{2014arXiv1412.0066G} all of our simulations, except FofR04, have a large degree of screening for our largest halo mass range. FofR04 does not have much screening even for the largest halos and consequently the modifications of gravity are active inside our most massive halos leading to a non-zero deviation even in the centre.

FofR04 is our simulation that is closest to act as a linear (non-screened) model. For this reason the modification in the temperature profiles is  close to a constant over the whole profile for both mass-ranges. This is to be expected since if we neglect the finite interaction range of the fifth-force then the modifications of gravity can be thought of as just a rescaling of the strength of gravity (i.e. of Newton's constant). For FofR04, the difference in amplitude in the temperature with respect to $\Lambda$CDM is seen to be slightly larger in the center for our smallest halos. This is likely related to the range of the fifth-force, being density dependent and smaller for denser objects, making the effect of the fifth-force (through not much screened) smaller for the largest halos.

For FofR05 we see a modification in the centre only for our smallest halos. This is also to be expected from Fig. 6  in \cite{2014arXiv1412.0066G} which shows that for FofR05 our large halos are very much screened, but our small halos are not much screened. For FofR06 we see only a very small (a few percent) deviation in the temperature for both of our halos mass-ranges over the whole profile again consistent with the findings of \cite{2014arXiv1412.0066G}. The FofR06 model has the highest amount of screening and generally the shortest range of all models and to see sizeable deviations in this model we would need to go to much smaller halo masses which is beyond the resolution of our simulations.

For the symmetron models we also see the effect of the additional symmetron screening mechanism. Sym D is the symmetron model that has the least amount of (non-linear) screening and the fifth-force has been in operation for the longest time. However, opposed to FofR04, the deviation in the centre of the largest halos is rather small. The reason for this is that the coupling to matter $\beta(\varphi) \propto \varphi$ is field-dependent and goes to zero for large and dense halos giving rise to additional screening which we don't have the $f(R)$ model where $\beta$ is a constant. This effect suppresses the modifications of gravity inside large halos even though we don't have a large degree of (non-linear) screening. For the other symmetron models (Sym A mostly) we also have a large degree of non-linear screening further suppressing the fifth-force and thereby suppressing the deviations seen in the temperature profiles.

The temperature profiles show all the different aspects of screening we have in the models we have simulated and because of this we would also expect to see an environment dependence of the temperature profiles for halos residing in different environments similar to what was found in \cite{Winther} for dark matter halos. This is left for future work.

\section{Power spectra}
\label{power_results}
The power spectra were computed using the publicly available POWMES code \citep{POWMES}, for both the dark matter and the gas. In order to use POWMES to compute the gas power spectrum we needed to extract the mass of each cell from the density of the cell. This is done by assigning a mass to grid cells using
\begin{align}
 m = \rho V_{\rm cell}.
\end{align}
where $\rho$ and $V_{\rm cell}$ correspond to the gas density and volume of each cell.

\begin{figure}
        \centering
        \includegraphics[width=0.48\textwidth]{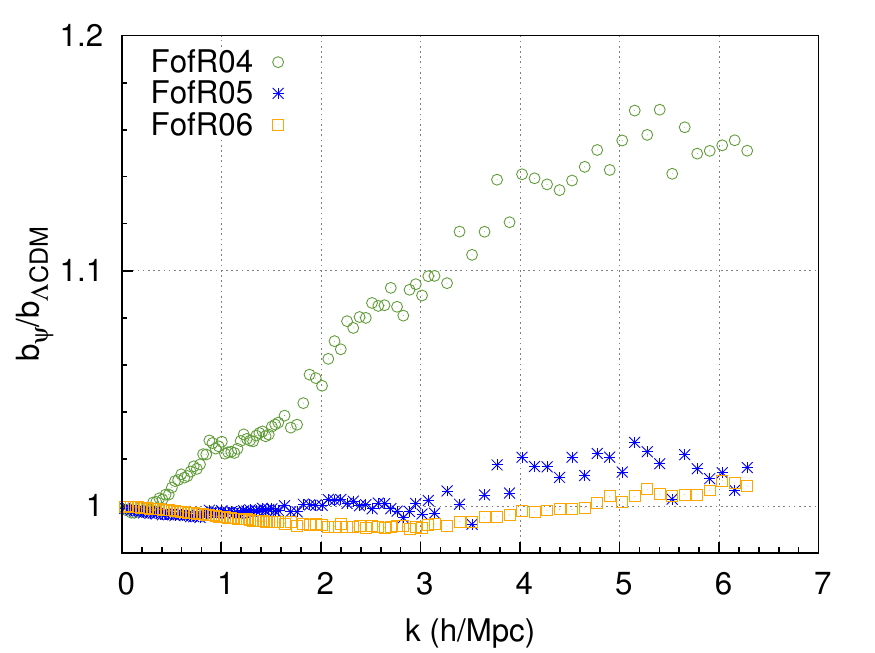}
\caption{\label{biasFoF} We show the the deviation of the bias from $\Lambda$CDM, for $f(R)$ gravity.}
\end{figure}
\begin{figure}
        \centering
        \includegraphics[width=0.48\textwidth]{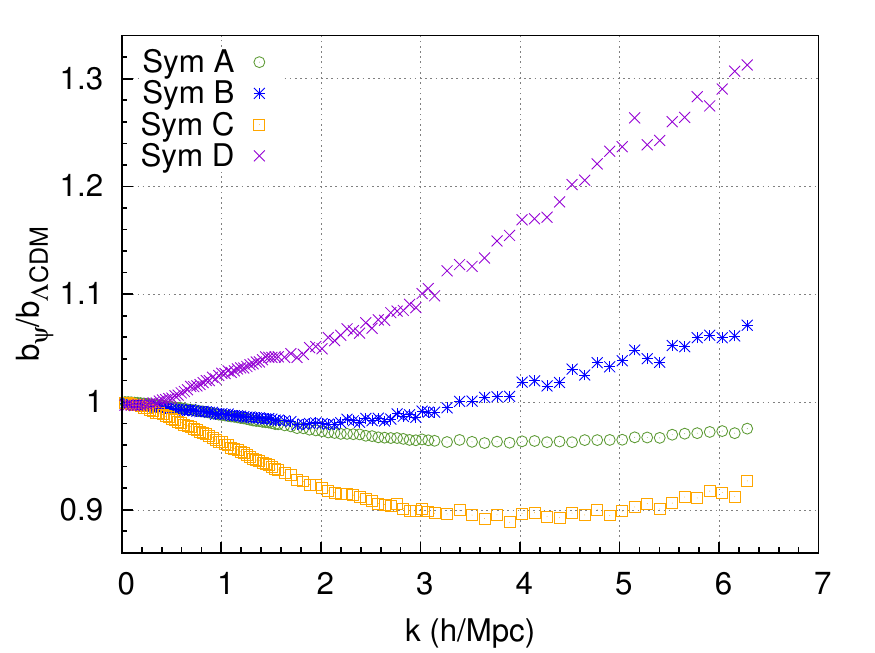}
\caption{\label{biasSym} We show the the deviation of the bias from $\Lambda$CDM for the symmetron model.}
\end{figure}

We will start by studying the effects modified gravity have on the ratio between the gas power spectrum and the dark matter power spectrum, 
\begin{align}
 b = \frac{P_{\rm DM}}{P_{\rm gas}}.
\end{align}
In \fig{biasFoF} and \fig{biasSym} we show the fractional difference of the gas-dark matter bias with respect to $\Lambda$CDM for both the $f(R)$ and symmetron simulations respectively. We first note how the bias is strongly scale dependent. At large scales ($k\lesssim 0.5$ h/Mpc) we see that all theories converge to the bias of $\Lambda$CDM. However once we move to smaller scales $k\gtrsim 0.5$ h/Mpc we see that the deviations of the bias grow at an increasing rate for most models.

 \begin{figure*}
         \centering
        \includegraphics[width=0.47\textwidth]{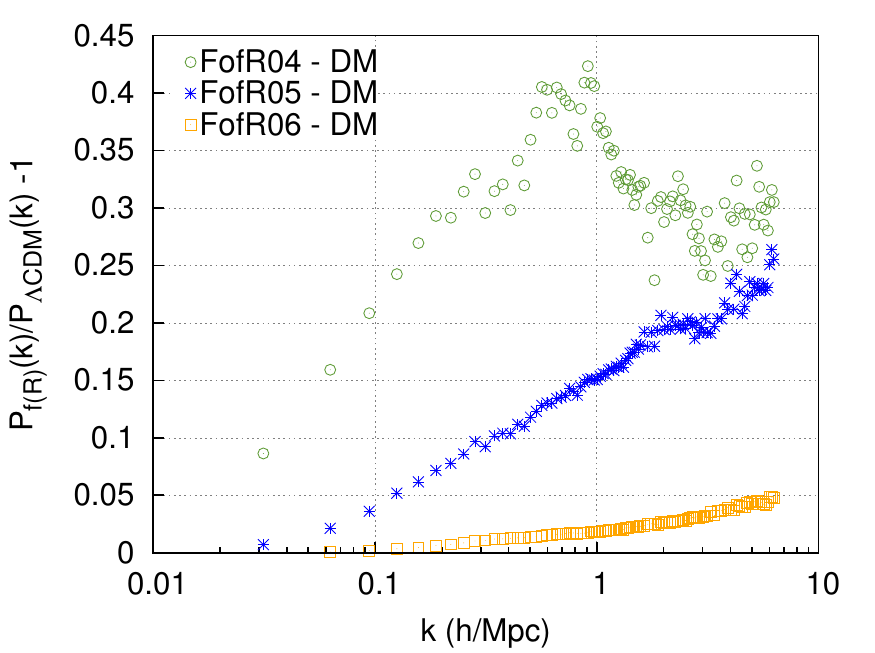}
	\vspace{-2 mm}
        \includegraphics[width=0.47\textwidth]{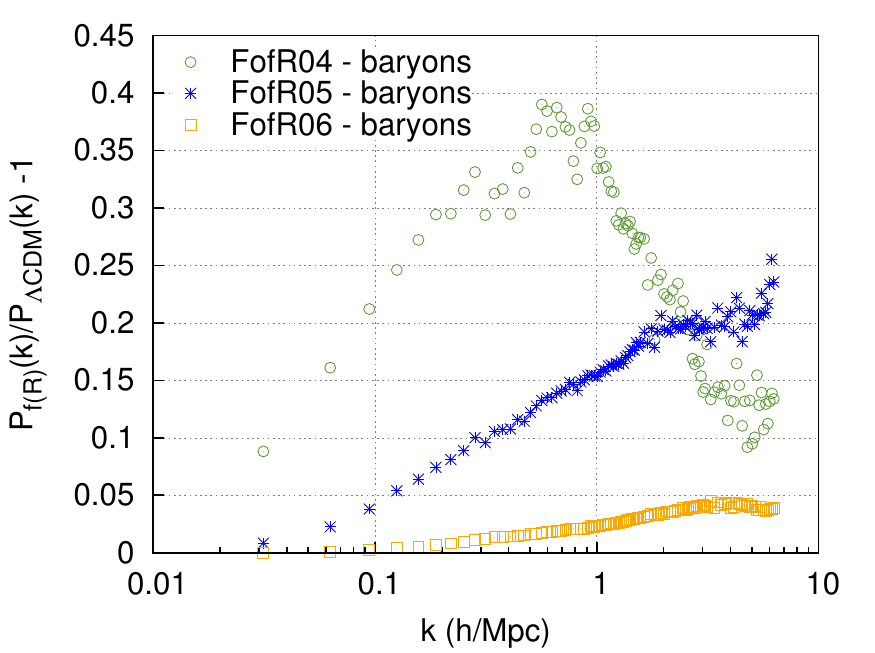}
	\vspace{-2 mm}
 \caption{The left and right figures show the power spectrum deviations from standard $\Lambda$CDM  for our various $f(R)$ models for CDM and baryons respectively.}\label{fig:FofR_Powerspectrums}
 \end{figure*}

 \begin{figure*}
         \centering
         \includegraphics[width=0.5\textwidth]{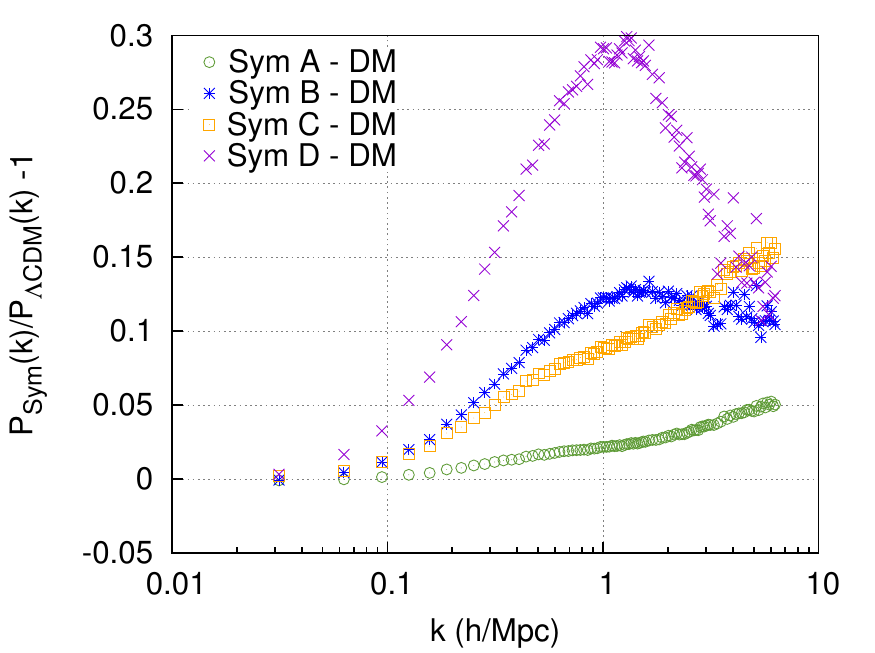}
 	\vspace{-2 mm}
 	\hspace{-7 mm}
         \includegraphics[width=0.5\textwidth]{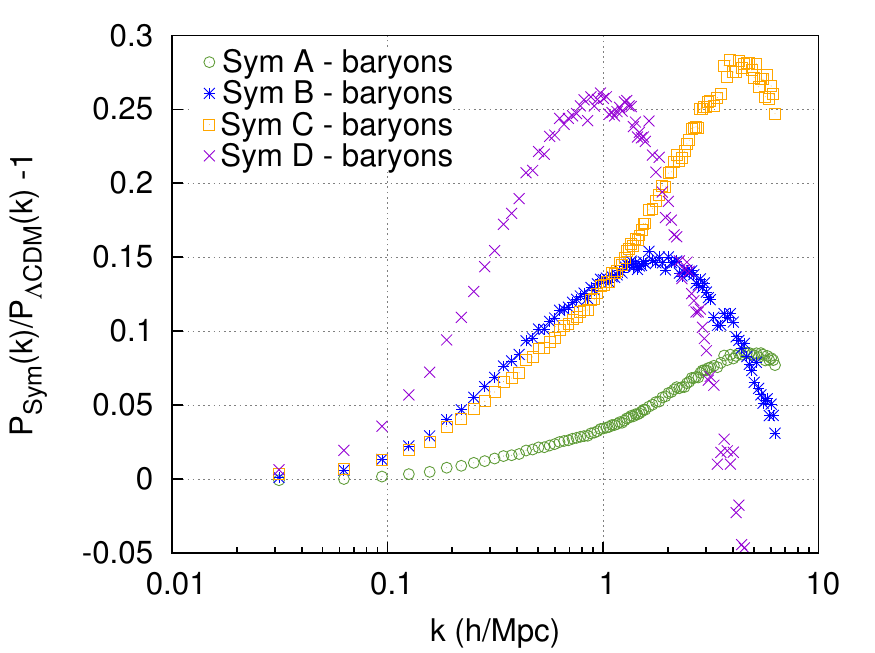}
	\vspace{-2 mm}
 \caption{The left and right figures show the power spectrum deviations from standard $\Lambda$CDM for our various symmetron models for CDM and baryons respectively.}\label{fig:Symmetron_Powerspectrums}
 \end{figure*}

Looking at Sym A, B and D (again, Sym C mirrors Sym A, but is more extreme) we see that the bias can both be higher and lower than the $\Lambda$CDM bias depending on $a_{\rm SSB}$. The bias seem to rotate in a clockwise motion as $a_{\rm SSB}$ increases. We also see a direct link between the strength of the coupling and the increased deviations in the bias when comparing Sym A and Sym C. For the $f(R)$ simulations we see that FofR04 has  a bias that increases strongly when we go to smaller scales, whereas FofR05 and FofR06 has a bias that is close to $\Lambda$CDM. It should be pointed out that Sym A and Sym B together with FofR05 and FofR06 are the most realistic models, whereas FofR04 and Sym D are extremes (which most likely can be ruled out already based on current data). The main thing we observe is that the behaviour of the bias can be very different depending on how long the fifth-force have been in operation, the range over which it acts, the coupling strength to matter and the amount of screening that is in play. This makes it hard to give a general prediction for what the gas does in these theories as it is very dependent on the model parameters. However, we do notice that it is in the simulations where we have a lower degree of screening (Sym D, FofR04 and to a lesser degree Sym B) that the relative bias grows significantly with scale. A possible explanation for this is that it is only the very dense gas inside clusters that experience much screening of the fifth-force and therefore do not cluster as expected.”

In \fig{fig:Symmetron_Powerspectrums} and \fig{fig:FofR_Powerspectrums} we present the deviations of the dark matter and gas power spectra for the symmetron models and the $f(R)$-gravity theories from $\Lambda$CDM respectively. For both models we see that the majority of the models peak at approximately 1 $h$/Mpc, which is the scale at which streaming velocities reach a value large enough to prevent further clustering. The results for dark matter is found to be in good agreement with existing simulations in the literature.

We start by noting a direct correlation between the amplitude of the power spectrum deviations and the time since symmetry breaking. The longer time the fifth-force has had a chance to work on the dark matter particles and the gas the larger deviations from $\Lambda$CDM we find, in agreement with expectations.

Further we see that the higher $a_{\rm SSB}$ we have, the lower the scale at which the screening mechanism starts, we might have expected a complete opposite effect as the density usually increases as we go to smaller and smaller scale, however this is clearly not what we observe. This might tell us that the screening mechanism is less effective for these values than naively expected.

By looking at Sym C we can see the sensitivity of the power spectrum with regards to the strength of the fifth-force. We observe that the effect of doubling the strength of the force compounds drastically in the non-linear regime, particularly for the gas component.

For the $f(R)$ case we know that in the FofR04 simulations we have much less screening than in the FofR05 and FofR06 simulations (as can be observed by comparing full simulations with linearised no-screening simulations, see \citet{Li1}). It is only in the most dense regions of the simulation box where the fifth-force is significantly screened for FofR04. As the most dense regions in the simulation box largely consists of highly clustered baryonic gas the gas will experience much more screening than the dark matter counterpart and this effect can explain the large growth of the bias with scale as we observe for FofR04. For models where the screening mechanism works effectively, like for FofR05 and FofR06 the fifth-force seem to affect the dark matter and gas equally well.

\section{Discussion}
\label{discuss}
In this paper we have studied the effects modified gravity has on structure formation with dark matter and a baryonic gas. The aim being to study the effects of modified gravity on the baryonic physics in its simplest form.

We ran several simulations with $256^3$ particles for the symmetron and $f(R)$-gravity models respectively. With the data from these simulations we analysed the density profiles and power spectra for both the dark matter and gas components, while also commenting on the behaviour of the gas-DM bias. 

We have shown that great care must be taken when trying to rule out modified gravity theories by comparing observations to the dark matter predictions, as the addition of the baryonic gas reduces the deviations from $\Lambda$CDM significantly, at least for density profiles. The smaller effect of modified gravity on the gas density profiles compared to the effect it has on the dark matter density profiles is most likely due to an environmental effect from the dark matter. Namely that the total density $\rho_{\rm tot} = \rho_{\rm DM}+\rho_{\rm gas}$ gets sufficiently high to trigger the screening mechanism before the gas component has had enough time to be acted upon by the extra force to give the expected impact. The collisionless nature of dark matter give rise to a lower density at the inner regions of the dark matter halos than the inner region density of the gas halos.

The analysis of power spectra shows us that it is mainly in the non-linear regime that the differences between the dark matter and the gas cases are significant, in agreement with previous findings. The power spectrum is highly sensitive to the range of the scalar field, and especially sensitive to the time that the fifth-force has had to affect the content of the Universe.

We have found that temperature profiles of clusters can be a strong signature of modified gravity. The deviation from the $\Lambda$CDM predictions can be a factor of a few larger than the same deviations found in the density profiles and power spectra, and much more sensitive to the fifth-force then previously assumed. It is also important to notice that the effects that modified gravity theories have on the baryonic gas does not always exactly mirror the effects it has on the dark matter.

Using these newfound characteristics of the symmetron and $f(R)$-gravity models of modified gravity on the temperature we can start comparing to probes of gravity such as in \citet{TestCham} or as suggested in \citet{novelprobe} and hopefully produce better constraints.

It is worth to note that our study have only focused on a simple baryonic gas without taking into account important physical effects we know take place in our Universe. In the future we aim to extend this work by adding such effects as feedback and cooling into our code. Feedback effects will particularly be interesting to see whether they enhance the deviations we already observe or not, whereas cooling might have significant effects on the extremities of the temperature profiles, where we observe the extreme deviations.

\section*{Acknowledgements}
The authors wish to thank Yuanchan Cai and Max B. Gr\"onke for useful discussions. AH, CL and DFM  would like to thank the Research Council of Norway for funding. HAW was supported by the BIPAC and the Oxford Martin School.

\bibliography{Nepht_bib}

\begin{thebibliography}{55}
\expandafter\ifx\csname natexlab\endcsname\relax\def\natexlab#1{#1}\fi

\bibitem[{{Arnold}, {Puchwein} \& {Springel}(2014){Arnold}, {Puchwein}, \&
  {Springel}}]{nrtwo}
{Arnold} C., {Puchwein} E., {Springel} V., 2014, \mnras, 440, 833

\bibitem[{Barreira {et~al}\mbox{.}(2013)Barreira, Li, Hellwing, Baugh, \&
  Pascoli}]{Barreira}
Barreira A., Li B., Hellwing W.~A., Baugh C.~M., Pascoli S., 2013, JCAP, 1310,
  027

\bibitem[{{Behroozi}, {Wechsler} \& {Wu}(2013){Behroozi}, {Wechsler}, \&
  {Wu}}]{rockstar}
{Behroozi} P.~S., {Wechsler} R.~H., {Wu} H.-Y., 2013, \apj, 762, 109

\bibitem[{{Bertotti}, {Iess} \& {Tortora}(2003){Bertotti}, {Iess}, \&
  {Tortora}}]{solarsyst}
{Bertotti} B., {Iess} L., {Tortora} P., 2003, \nat, 425, 374

\bibitem[{{Bertschinger}(1999)}]{Grafic}
{Bertschinger} E., 1999, {COSMICS: Cosmological initial conditions and
  microwave anisotropy codes}. Astrophysics Source Code Library

\bibitem[{{Brax} {et~al}\mbox{.}(2012){Brax}, {Davis}, {Li}, \&
  {Winther}}]{screen2}
{Brax} P., {Davis} A.-C., {Li} B., {Winther} H.~A., 2012, \prd, 86, 044015

\bibitem[{{Brax} {et~al}\mbox{.}(2013){Brax}, {Davis}, {Li}, {Winther}, \&
  {Zhao}}]{nbodycham}
{Brax} P., {Davis} A.-C., {Li} B., {Winther} H.~A., {Zhao} G.-B., 2013, \jcap,
  4, 29

\bibitem[{Brookfield {et~al}\mbox{.}(2006)Brookfield, van~de Bruck, Mota, \&
  Tocchini-Valentini}]{brook}
Brookfield A.~W., van~de Bruck C., Mota D., Tocchini-Valentini D., 2006,
  Phys.Rev., D73, 083515

\bibitem[{{Brown} \& {Hammami}(2012)}]{BrownHammami}
{Brown} I.~A., {Hammami} A., 2012, \jcap, 4, 2

\bibitem[{{Capozziello} \& {de Laurentis}(2011)}]{Capozziello}
{Capozziello} S., {de Laurentis} M., 2011, \physrep, 509, 167

\bibitem[{{Cen}(1992)}]{MethodologyPaper}
{Cen} R., 1992, \apjs, 78, 341

\bibitem[{{Colombi} \& {Novikov}(2011)}]{POWMES}
{Colombi} S., {Novikov} D., 2011, {POWMES: Measuring the Power Spectrum in an
  N-body Simulation}. Astrophysics Source Code Library

\bibitem[{{Copeland}, {Sami} \& {Tsujikawa}(2006){Copeland}, {Sami}, \&
  {Tsujikawa}}]{darkenergy}
{Copeland} E.~J., {Sami} M., {Tsujikawa} S., 2006, International Journal of
  Modern Physics D, 15, 1753

\bibitem[{{de Felice} \& {Tsujikawa}(2010)}]{f(R)Theories}
{de Felice} A., {Tsujikawa} S., 2010, Living Reviews in Relativity, 13, 3

\bibitem[{{Dimopoulos} {et~al}\mbox{.}(2007){Dimopoulos}, {Graham}, {Hogan}, \&
  {Kasevich}}]{labexp2}
{Dimopoulos} S., {Graham} P.~W., {Hogan} J.~M., {Kasevich} M.~A., 2007,
  Physical Review Letters, 98, 111102

\bibitem[{Faraoni, Gunzig \& Nardone(1999)Faraoni, Gunzig, \&
  Nardone}]{Faraoni:1998qx}
Faraoni V., Gunzig E., Nardone P., 1999, \fcp, 20, 121

\bibitem[{{Fujii} \& {Maeda}(2003)}]{2003CQGra..20.4503F}
{Fujii} Y., {Maeda} K.-i., 2003, Classical and Quantum Gravity, 20, 4503

\bibitem[{{Gronke} {et~al}\mbox{.}(2014){Gronke}, {Llinares}, {Mota}, \&
  {Winther}}]{2014arXiv1412.0066G}
{Gronke} M., {Llinares} C., {Mota} D.~F., {Winther} H.~A., 2014, ArXiv e-prints

\bibitem[{{Gronke}, {Llinares} \& {Mota}(2014){Gronke}, {Llinares}, \&
  {Mota}}]{symmfofrredshift}
{Gronke} M.~B., {Llinares} C., {Mota} D.~F., 2014, \aap, 562, A9

\bibitem[{Hinterbichler \& Khoury(2010)}]{SymmetronPaper}
Hinterbichler K., Khoury J., 2010, \prl, 104, 231301

\bibitem[{{Hoyle} {et~al}\mbox{.}(2004){Hoyle}, {Kapner}, {Heckel},
  {Adelberger}, {Gundlach}, {Schmidt}, \& {Swanson}}]{labexp}
{Hoyle} C.~D., {Kapner} D.~J., {Heckel} B.~R., {Adelberger} E.~G., {Gundlach}
  J.~H., {Schmidt} U., {Swanson} H.~E., 2004, \prd, 70, 042004

\bibitem[{{Hu} \& {Sawicki}(2007)}]{husawicki}
{Hu} W., {Sawicki} I., 2007, \prd, 76, 064004

\bibitem[{{Jain} {et~al}\mbox{.}(2013){Jain}, {Gabadadze}, {Hu}, {Hui},
  {Huterer}, {Kamionkowski}, {Khoury}, {Koyama}, {Li}, {Linder}, {Schmidt},
  {Scoccimarro}, {Starkman}, {Stubbs}, {Takada}, {Tolley}, {Trodden}, {Uzan},
  {Vikram}, {Weltman}, {Wyman}, {Zaritsky}, \& {Zhao}}]{novelprobe}
{Jain} B. {et~al.}, 2013, ArXiv e-prints

\bibitem[{{Khoury}(2010)}]{screen1}
{Khoury} J., 2010, ArXiv e-prints

\bibitem[{{Khoury} \& {Weltman}(2004)}]{Chameleons}
{Khoury} J., {Weltman} A., 2004, \prd, 69, 044026

\bibitem[{{Kroupa} {et~al}\mbox{.}(2010){Kroupa}, {Famaey}, {de Boer},
  {Dabringhausen}, {Pawlowski}, {Boily}, {Jerjen}, {Forbes}, {Hensler}, \&
  {Metz}}]{LCDMprob1}
{Kroupa} P. {et~al.}, 2010, \aap, 523, A32

\bibitem[{{Li} {et~al}\mbox{.}(2013){Li}, {Hellwing}, {Koyama}, {Zhao},
  {Jennings}, \& {Baugh}}]{2013MNRAS.428..743L}
{Li} B., {Hellwing} W.~A., {Koyama} K., {Zhao} G.-B., {Jennings} E., {Baugh}
  C.~M., 2013, \mnras, 428, 743

\bibitem[{Li, Mota \& Barrow(2011)Li, Mota, \& Barrow}]{Li1}
Li B., Mota D.~F., Barrow J.~D., 2011, Astrophys.J., 728, 109

\bibitem[{{Li}, {Zhao} \& {Koyama}(2013){Li}, {Zhao}, \&
  {Koyama}}]{nbodyvainst}
{Li} B., {Zhao} G.-B., {Koyama} K., 2013, \jcap, 5, 23

\bibitem[{{Li} {et~al}\mbox{.}(2012){Li}, {Zhao}, {Teyssier}, \&
  {Koyama}}]{Li2}
{Li} B., {Zhao} G.-B., {Teyssier} R., {Koyama} K., 2012, \jcap, 1, 51

\bibitem[{{Llinares}, {Knebe} \& {Zhao}(2008){Llinares}, {Knebe}, \&
  {Zhao}}]{2008arXiv0809.2899L}
{Llinares} C., {Knebe} A., {Zhao} H., 2008, \mnras, 391, 1778

\bibitem[{{Llinares} \& {Mota}(2013)}]{2013PhRvL.110p1101L}
{Llinares} C., {Mota} D.~F., 2013, Physical Review Letters, 110, 161101

\bibitem[{{Llinares} \& {Mota}(2014)}]{nbodysymmnqs}
{Llinares} C., {Mota} D.~F., 2014, \prd, 89, 084023

\bibitem[{{Llinares}, {Mota} \& {Winther}(2014){Llinares}, {Mota}, \&
  {Winther}}]{ISIS}
{Llinares} C., {Mota} D.~F., {Winther} H.~A., 2014, \aap, 562, A78

\bibitem[{{Lombriser} {et~al}\mbox{.}(2013){Lombriser}, {Li}, {Koyama}, \&
  {Zhao}}]{BajiLi}
{Lombriser} L., {Li} B., {Koyama} K., {Zhao} G.-B., 2013, \prd, 87, 123511

\bibitem[{{Martel} \& {Shapiro}(1998)}]{SuperCom}
{Martel} H., {Shapiro} P.~R., 1998, \mnras, 297, 467

\bibitem[{{Misner}, {Thorne} \& {Wheeler}(1973){Misner}, {Thorne}, \&
  {Wheeler}}]{Gravitation}
{Misner} C.~W., {Thorne} K.~S., {Wheeler} J.~A., 1973, {Gravitation}

\bibitem[{{Mota} \& {Shaw}(2007)}]{motashaw}
{Mota} D.~F., {Shaw} D.~J., 2007, \prd, 75, 063501

\bibitem[{{Neto} {et~al}\mbox{.}(2007){Neto}, {Gao}, {Bett}, {Cole}, {Navarro},
  {Frenk}, {White}, {Springel}, \& {Jenkins}}]{relax}
{Neto} A.~F. {et~al.}, 2007, \mnras, 381, 1450

\bibitem[{{Noller}, {von Braun-Bates} \& {Ferreira}(2014){Noller}, {von
  Braun-Bates}, \& {Ferreira}}]{quasistatic}
{Noller} J., {von Braun-Bates} F., {Ferreira} P.~G., 2014, \prd, 89, 023521

\bibitem[{{Oyaizu}(2008)}]{fofrnbodychicago}
{Oyaizu} H., 2008, \prd, 78, 123523

\bibitem[{{Peebles} \& {Ratra}(2003)}]{LCDMprob2}
{Peebles} P.~J., {Ratra} B., 2003, Reviews of Modern Physics, 75, 559

\bibitem[{{Puchwein}, {Baldi} \& {Springel}(2013){Puchwein}, {Baldi}, \&
  {Springel}}]{MG-Gadget}
{Puchwein} E., {Baldi} M., {Springel} V., 2013, \mnras, 436, 348

\bibitem[{{Ryu} {et~al}\mbox{.}(1993){Ryu}, {Ostriker}, {Kang}, \&
  {Cen}}]{RyOo}
{Ryu} D., {Ostriker} J.~P., {Kang} H., {Cen} R., 1993, \apj, 414, 1

\bibitem[{{Schmidt}(2009)}]{nbodydgp}
{Schmidt} F., 2009, \prd, 80, 123003

\bibitem[{{Shaw} {et~al}\mbox{.}(2006){Shaw}, {Weller}, {Ostriker}, \&
  {Bode}}]{Shaw}
{Shaw} L.~D., {Weller} J., {Ostriker} J.~P., {Bode} P., 2006, \apj, 646, 815

\bibitem[{{Sotiriou}(2006)}]{Sotiriou:2006hs}
{Sotiriou} T.~P., 2006, Classical and Quantum Gravity, 23, 5117

\bibitem[{{Springel}(2005)}]{Gadget}
{Springel} V., 2005, \mnras, 364, 1105

\bibitem[{{Terukina} {et~al}\mbox{.}(2014){Terukina}, {Lombriser}, {Yamamoto},
  {Bacon}, {Koyama}, \& {Nichol}}]{TestCham}
{Terukina} A., {Lombriser} L., {Yamamoto} K., {Bacon} D., {Koyama} K., {Nichol}
  R.~C., 2014, \jcap, 4, 13

\bibitem[{{Teyssier}(2002)}]{Ramses}
{Teyssier} R., 2002, \aap, 385, 337

\bibitem[{{Turk} {et~al}\mbox{.}(2011){Turk}, {Smith}, {Oishi}, {Skory},
  {Skillman}, {Abel}, \& {Norman}}]{YT-Project}
{Turk} M.~J., {Smith} B.~D., {Oishi} J.~S., {Skory} S., {Skillman} S.~W.,
  {Abel} T., {Norman} M.~L., 2011, \apjs, 192, 9

\bibitem[{{Vogelsberger} {et~al}\mbox{.}(2014){Vogelsberger}, {Genel},
  {Springel}, {Torrey}, {Sijacki}, {Xu}, {Snyder}, {Bird}, {Nelson}, \&
  {Hernquist}}]{Illustris}
{Vogelsberger} M. {et~al.}, 2014, \nat, 509, 177

\bibitem[{{Will}(2014)}]{cwillreview}
{Will} C.~M., 2014, Living Reviews in Relativity, 17, 4

\bibitem[{{Winther}, {Mota} \& {Li}(2012){Winther}, {Mota}, \& {Li}}]{Winther}
{Winther} H.~A., {Mota} D.~F., {Li} B., 2012, \apj, 756, 166

\bibitem[{{Zhao}, {Li} \& {Koyama}(2011){Zhao}, {Li}, \& {Koyama}}]{Li3}
{Zhao} G.-B., {Li} B., {Koyama} K., 2011, \prd, 83, 044007

\end{thebibliography}

\end{document}